\documentclass[letterpaper]{article}

\usepackage{arxiv}

\usepackage{amsmath}
\usepackage{amsfonts}
\usepackage{amssymb}

\usepackage[utf8]{inputenc} 
\usepackage[T1]{fontenc}    
\usepackage[colorlinks=true,linkcolor=blue,citecolor=blue,urlcolor=blue]{hyperref}
\usepackage[nameinlink,capitalize]{cleveref}
\usepackage{url}            
\usepackage{booktabs}       

\usepackage{nicefrac}       
\usepackage{microtype}      
\usepackage{graphicx}
\usepackage{caption}
\usepackage[numbers]{natbib}
\usepackage{doi}

\usepackage[section]{placeins} 

\usepackage[dvipsnames]{xcolor}
\usepackage{makecell}
\usepackage[version=4]{mhchem}
\usepackage[per-mode=symbol,round-precision=3,scientific-notation=false,range-phrase = \text{--}]{siunitx}

\usepackage{floatrow}
\usepackage{float}
\floatstyle{plaintop}
\restylefloat{table}

\usepackage{framed} 
\usepackage{multicol} 
\usepackage{nomencl} 
\usepackage{ifthen}
\renewcommand\nomgroup[1]{%
  \ifthenelse{\equal{#1}{A}}{%
    \item[\textbf{Acronyms}]}{
  \ifthenelse{\equal{#1}{R}}{
    \item[\textbf{Roman Symbols}]}{
  \ifthenelse{\equal{#1}{G}}{%
    \item[\textbf{Greek Symbols}]}{
  \ifthenelse{\equal{#1}{S}}{%
    \item[\textbf{Superscripts}]}{
  \ifthenelse{\equal{#1}{U}}{%
    \item[\textbf{Subscripts}]}{
  \ifthenelse{\equal{#1}{X}}{%
    \item[\textbf{Other Symbols}]}{
  {}}}}}}}}
\renewcommand*{\nompreamble}{\markboth{\nomname}{\nomname}}

\makenomenclature
\renewcommand*\nompreamble{\begin{multicols}{2}}
\renewcommand*\nompostamble{\end{multicols}}

\newcommand\T{\rule{0pt}{2.6ex}}
\newcommand\B{\rule[-1.2ex]{0pt}{0pt}}

\usepackage{xspace}
\newcommand*{\eg}{e.g.,\@\xspace}
\newcommand*{\ie}{i.e.,\@\xspace}

\newcommand*{\vs}{vs.\@\xspace}

\makeatletter
\newcommand*{\etc}{%
    \@ifnextchar{.}%
        {etc}%
        {etc.\@\xspace}%
}
\makeatother

\DeclareUnicodeCharacter{025B}{\ensuremath{\varepsilon}}

\graphicspath{{figures/}}

\usepackage{listings}

\definecolor{bkgd}{RGB}{240,242,246}
\definecolor{ceruleanblue}{rgb}{0.16, 0.32, 0.75}
\definecolor{orange-red}{rgb}{1.0, 0.27, 0.0}
\definecolor{anotherblue}{RGB}{37,92,243}
\definecolor{blackblue}{RGB}{46,60,85}
\definecolor{goldyellow}{RGB}{199,146,12}
\lstdefinestyle{altstyle2}{
    backgroundcolor=\color{bkgd},
    basicstyle=\ttfamily\footnotesize\color{blackblue},
    breakatwhitespace=false,
    breaklines=true,
    captionpos=b,
    commentstyle=\color{goldyellow},
    keepspaces=true,
    keywordstyle=\color{orange-red},
    language=Python,
    numbersep=5pt,
    numberstyle=\tiny\color{ceruleanblue},
    showspaces=false,
    showstringspaces=false,
    showtabs=false,
    stringstyle=\color{anotherblue},
    tabsize=2,
    numbers=left
}
\newcommand{\mtc}[1]{\makecell[tc]{#1}}

\newcommand{\mtl}[1]{\makecell[tl]{#1}}

\newcommand{\pkg}[1]{\texttt{#1}}

\renewcommand{\O}[1]{$\mathcal{O}(#1)$}

\usepackage[scaled=0.86]{helvet}

\crefname{lstlisting}{listing}{listings}
\Crefname{lstlisting}{Listing}{Listings}

\usepackage{pdflscape}
\newcommand{\nusvr}{$\nu{\text -}\mathrm{SVR}$\@\xspace}

\DeclareMathOperator*{\argmin}{arg\,min}

\title{Empirical Models for Multidimensional Regression of Fission Systems}

\author{ 
	Akshay J.~Dave \\
	Nuclear Reactor Laboratory\\
	Massachusetts Institute of Technology\\
	Cambridge, MA 02139 \\
	\texttt{akshayjd@mit.edu} \\
	\And
	Jiankai Yu \\
	Nuclear Science and Engineering\\
	Massachusetts Institute of Technology\\
	Cambridge, MA 02139 \\
	\And
	Jarod Wilson \\
	Nuclear Reactor Laboratory\\
	Massachusetts Institute of Technology\\
	Cambridge, MA 02139 \\
	\And
	Bren Phillips \\
	Nuclear Science and Engineering\\
	Massachusetts Institute of Technology\\
	Cambridge, MA 02139 \\
	\And
	Kaichao Sun \\
	Nuclear Reactor Laboratory\\
	Massachusetts Institute of Technology\\
	Cambridge, MA 02139 \\
	\And
	Benoit Forget \\
	Nuclear Science and Engineering\\
	Massachusetts Institute of Technology\\
	Cambridge, MA 02139 \\
}

\date{}

\renewcommand{\headeright}{ }
\renewcommand{\undertitle}{ }
\renewcommand{\shorttitle}{Empirical models for multidimensional regression of fission systems}

\begin{document}
\maketitle

\begin{abstract}
The development of next-generation autonomous control of fission systems, such as nuclear power plants, will require leveraging advancements in machine learning.
For fission systems, accurate prediction of nuclear transport is important to quantify the safety margin and optimize performance.
The state-of-the-art approach to this problem is costly Monte Carlo (MC) simulations to approximate solutions of the neutron transport equation.
Such an approach is feasible for offline calculations \eg for design or licensing, but is precluded from use as a model-based controller.
In this work, we explore the use of Artificial Neural Networks (ANN), Gradient Boosting Regression (GBR), Gaussian Process Regression (GPR) and Support Vector Regression (SVR) to generate empirical models.
The empirical model can then be deployed, \eg in a model predictive controller.
Two fission systems are explored: the subcritical MIT Graphite Exponential Pile (MGEP), and the critical MIT Research Reactor (MITR).
A meta-learning approach is adopted to optimize each combination of machine learning algorithm and physical system.

\vspace{6pt}
Findings from this work establish guidelines for developing empirical models for multidimensional regression of neutron transport.
Our work finds that the qualitative differences in both fission systems manifest in different optimal hyperparameter sets.
An assessment of the accuracy and precision finds that the SVR, followed closely by ANN, performs the best.
For both MGEP and MITR, the optimized SVR model exhibited a domain-averaged, test, mean absolute percentage error of 0.17 \%.
A spatial distribution of performance metrics indicates that physical regions of poor performance coincide with locations of largest neutron flux perturbation -- this outcome is mitigated by ANN and SVR.
Even at local maxima, ANN and SVR bias is within experimental uncertainty bounds.
A comparison of the performance \vs training dataset size found that SVR is more data-efficient than ANN.
Both ANN and SVR achieve a greater than 7 order reduction in evaluation time \vs a MC simulation.
The GBR algorithm was not recommended due to its large generalization error.
The GPR algorithm had moderate performance, yet provides a path to quantify posterior uncertainty distribution.
\end{abstract}


\cleardoublepage
\section{Introduction}

The development of semi or fully autonomous Nuclear Power Plant (NPP) systems will require embedded surrogate models that can compress expensive calculations, while maximizing accuracy, precision, and the capability to generalize.
While several hydraulic components of the entire plant may be accurately represented by 1-D transport equations, the nuclear core requires greater fidelity.
Modeling the nuclear reactor core hinges on the capability to simulate neutron transport.
The steady-state integro-differential neutron transport equation is \citep[Ch. 4.II]{duderstadt1976nuclear},
\begin{equation} \label{eq:transport}
\mathbf{\Omega}\cdot\mathbf{\nabla}\phi(\mathbf{x},E,\mathbf{\Omega})+\Sigma_t(E)\phi(\mathbf{x},E,\mathbf{\Omega})=
\int_{4\pi}d\mathbf{\Omega}'\int_0^\infty dE'\Sigma_s(E'\rightarrow E,\mathbf{\Omega}'\rightarrow\mathbf{\Omega})\phi(\mathbf{x},E,\mathbf{\Omega})+s(\mathbf{x},E,\mathbf{\Omega})~,
\end{equation}
where $\phi$ is the angular neutron flux, $\Sigma_t$ and $\Sigma_s$ are the neutron total and scattering macroscopic cross sections, and $s$ is the neutron source term.
There are several independent variables: $\mathbf{x}$, a 3-D vector representing space; $\mathbf{\Omega}$ a 2-D vector representing unit angular direction of travel; and energy, $E$.
Therefore, for steady-state problems, $\phi$ is a function of \textit{six} independent variables.
The macroscopic cross sections ($\Sigma$) represent various interactions between neutrons and matter.
The interactions are a function of material type and energy (the latter is specifically challenging due to the presence of nonlinear resonance regions).
The units and interpretation of each term of \cref{eq:transport} is provided in \cref{appendix:A}.
In brief: \cref{eq:transport} accounts for the population balance of neutrons in 6-D phase space. 
There are approximate solutions for transport in idealized conditions (\eg homogenization, symmetry, energy-independence, and angular-independence).
However, exact solutions of \cref{eq:transport} are not feasible for realistic systems with complex geometries.
There are two paths to approximating high-fidelity solutions: deterministic and Monte Carlo \citep[Ch. 5.10]{cacuci2010handbook}.
Deterministic methods, discretize and solve \cref{eq:transport}, resulting in extraordinarily large system of equations (\eg \O{10^9} unknowns for a 100 grid 3-D system with 20 energy groups and 50 directions of flight).
Monte Carlo (MC) methods model nuclear transport using probabilistic methods, simulating a neutron's life within the system.
Given sufficiently large histories, the results from MC are statistically stationary and approximates \cref{eq:transport}.
Updates to $\Sigma$ are necessary over time (\eg as fuel is depleted), but this is handled by an external depletion solver.
MC is more widely used due to its ease of implementation, ability to treat complex geometries, and address issues arising from irregular $\Sigma$.
However, MC methods are still computationally intensive, requiring \O{10^7} s total wall-time for a full-core simulation of a small reactor (\eg MITR described in \cref{subsec:systems}).

To improve cost effectiveness of NPP operation, autonomous control is an upcoming area of research.
Within proposed autonomous frameworks \citep{Wood2017}, there is a need for surrogate models to provide diagnosis and prognosis of irregular operation, sensor-failure, or long-term performance degradation (such as heat-exchanger fouling).
Most ex-core components of the NPP can be characterized by linearizable differential equations (\eg fluid transport through pipes, or heat transfer between components).
However, in order to model the core itself, the solution to \cref{eq:transport} is required.
As highlighted above, simply embedding entire fundamental models is not feasible due to the high compute requirement.
An option is to create data-driven empirical models using machine learning (ML).
Developing high-fidelity empirical models allows us to develop model predictive controllers that can address problems with greater granularity \textit{during} NPP operation, \eg optimizing fuel depletion with asymmetric $\phi$ manipulation or increasing core safety margins with multidimensional predictions of power distribution.

\subsection{Previous Work}

The application of ML towards NPP oriented problems has been ongoing since the 90s.
The previous work can be classified into several major categories.
We will highlight previous work, and detail those that are most relevant to our topic of multidimensional neutron transport regression.
The papers utilize forms of Artificial Neural Networks (ANN), Gaussian Process Regression (GPR) or Support Vector Regression (SVR).
No published work was found that utilized Gradient Boosting Regression (GBR), an algorithm explored in this study.

Previous work is categorized into several groups.
Several authors have focused on \textit{classification of transients}, where the objective is to diagnose the occurrence of anomalous events using plant parameters as input.
ANNs were used to classify transients \citep{Basu1994,Santosh2007,Bartlett2017} such as: hot- and cold-leg loss of coolant, control rod ejection, total loss of off-site power, main steamline break, main feedwater line break and steam generator tube leak accidents.
A majority of studies focused on regression problems, which can be grouped into several subcategories.
Several authors have focused on \textit{non-operational parameter regression}, where the objective is to optimize \eg the fuel loading pattern using ANNs \citep{Sadighi2002,Leniau2015} or reinforcement learning \citep{radaideh2021physics}.
A plurality of work has focused on \textit{operational parameter regression}, where 0-D regression of plant parameters during operation was sought.
ANNs were used to model the reactor thermal power, \citep{Roh1991,Roh1991a}, the critical heat flux \citep{Kim1997,Zhao2020}, power-peaking factors \citep{Souza2006,Mirvakili2012}, and criticality swings due to fuel burn-up \citep{Mazrou2009}.
GPR has been used to predict component degradation \cite{baraldi2015prognostics,lee2019enhanced}.
SVR has been used to model 0-D \citep{lee2010prediction,liu2013nuclear} and 1-D critical heat flux \citep{kim2011prediction}, and power-peaking factors \cite{bae2008calculation}.
Recurrent networks for \textit{time-series regression} of point parameters has also been explored \citep{Nguyen2020,wang2021advanced}.

The category of \textit{multidimensional regression} was scarcely explored.
Only a single study was found: \citet{Boroushaki2005} used a so-called cellular neural network to predict the transient 3-D power distribution of a theoretical homogeneous cubic reactor. 
The authors reported a total percentage error (integrated over the entire transient) of 5.9\%.
In order to address this deficit, our preliminary work focused on establishing the performance of ANNs in generating multidimensional models for an actual nuclear system, and using modern libraries \citep{dave2020deep}.
We found that optimized ANNs were able to generalize well, and, across all \textit{test} datasets, achieved a mean absolute percentage error of 1.16 \% with a corresponding standard deviation of 0.77 \%.
The literature review indicated that GPR, SVR and GBR has not been explored for multidimensional regression of nuclear fission systems.
Furthermore, we also found that most studies focus on the application of one (or occasionally two) ML algorithms to a single nuclear system.
The application of the same ML algorithm to more than one system was not explored in the literature.

\subsection{Systems Modeled}\label{subsec:systems}

\begin{figure}
\includegraphics[width=\linewidth]{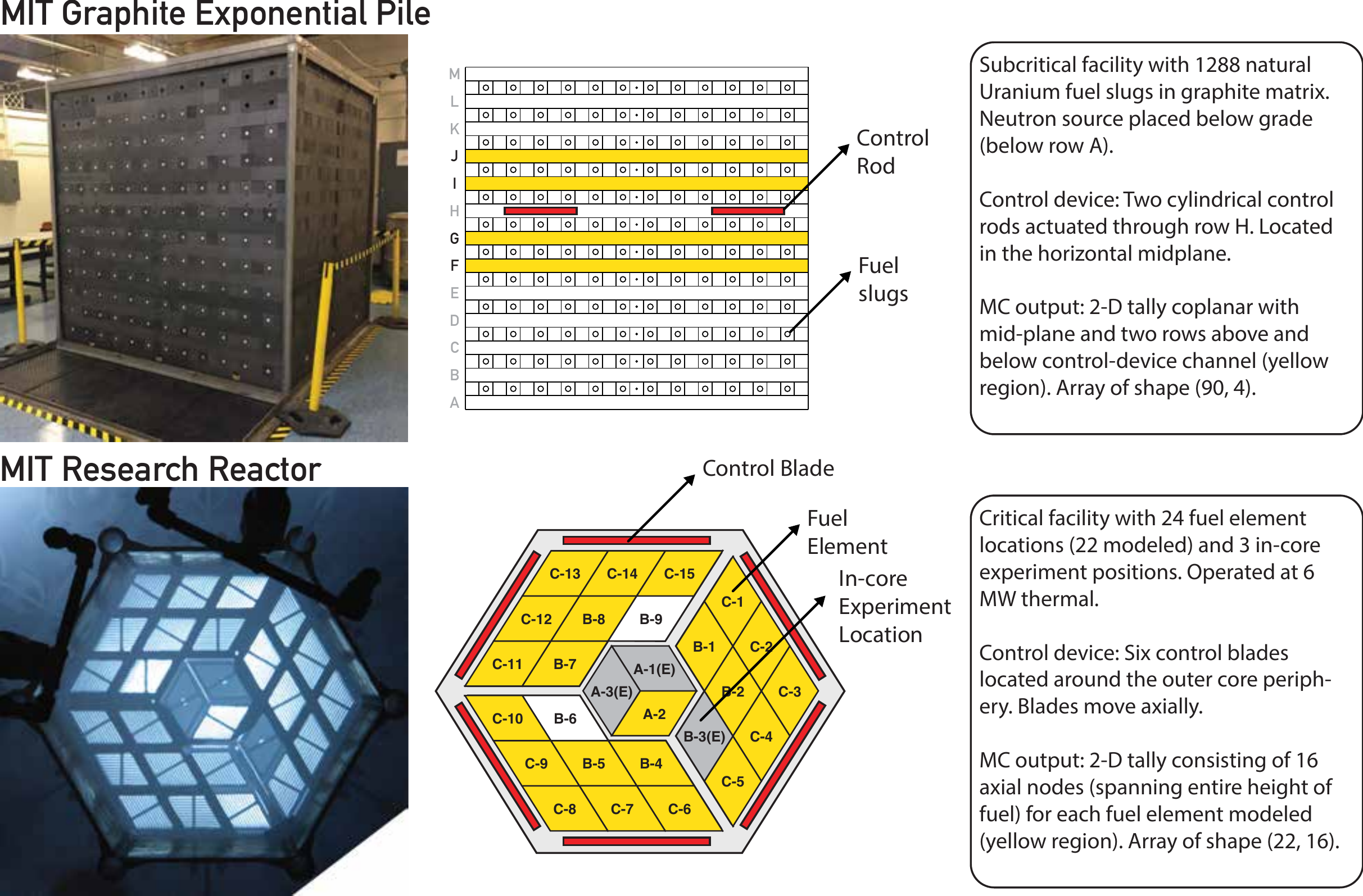}
\caption{Overview of facilities modeled in this work. Top row: the MGEP (only the fueled above-grade section is shown). Bottom row: the MITR. Inset is a description of the facilities, control devices and expected MC output shape. For both facilities' empirical model: the input is the position of the control devices, and the output is MC solutions of the neutron transport equation.}
\label{fig:systems}
\end{figure}

In this work, we consider creating empirical models of two nuclear systems.
The first is a subcritical nuclear system -- the MIT Graphite Exponential Pile (MGEP).
The second is a critical nuclear system -- the MIT Research Reactor (MITR).
Both systems are interesting to explore as they have qualitative differences in their response to external perturbation.
The differences are outlined in \cref{appendix:A}, where solutions to the neutron transport equation for canonical subcritical and critical systems are derived.
An overview of the facilities modeled is presented in \cref{fig:systems}.

The MGEP facility was developed at MIT in the 1950s.
The facility is made of two parts: an above-grade 90 inch cubic graphite lattice, and a below-grade graphite pedestal.
The above-grade part is fueled and, when fully loaded, contains 1288 natural uranium fuel slugs.
The pedestal consists of a graphite lattice that is used for insertion of neutron sources (either \ce{PuBe} or \ce{^{252}Cf}).
Records indicate that from the 1970s-2010s, the facility was rarely used and knowledge of its existence was uncommon.
The facility was recently re-started in 2016 and its components thoroughly characterized \cite{Gale2018}.
The graphite pile is subcritical, which means that the number of neutrons created by fission is lower than those absorbed, \ie the neutron population is not self-sustaining.
Therefore, in order to have a steady-state neutron distribution within the facility, an external neutron source is required.
As the facility remains subcritical when any additional absorbing media is inserted, it is an ideal testbed for conducting experiments.
In the past, the MGEP was primarily used for educational purposes.
Our ongoing research aims to transform the MGEP into a testbed for developing ML controllers.
In particular, we have developed an in-pile facility to actuate control rods and neutron detectors.
Through these efforts, we will be able to simultaneously perturb and monitor the time-dependent neutron flux distribution.

The MGEP nomenclature used to identify accessible layers is shown in \cref{fig:systems}.
The placement of the control rods in layer H is shown.
Control rods are made of a strong neutron absorber (\ie materials with large neutron absorption cross-section); in general, different forms of Boron and Cadmium are used for research reactors.
No control rods had existed for the MGEP.
Thus, we fabricated rods from scratch by filling aluminum pipes with a Boron-epoxy.
The custom rods significantly depressesed the local neutron flux ($\approx20\%$ perturbation in experimental tests).
The highlighted F, G, I and J layers, indicate potential locations for flux monitoring, and thus neutron detector placement.
The efforts to procure experimental distributions of neutron flux with varying control rod locations is ongoing.
Meanwhile, we have developed an \pkg{OpenMC} \cite{romano2013openmc} model to obtain numerical approximations of \cref{eq:transport}.
\pkg{OpenMC} is an open-source code, with development initiated at the Massachusetts Institute of Technology.
The entire 3-D structure of the MGEP is modeled.
However, only 2-D tallies in the regions for potential detector placement are collected.
Each layer is discretized into 90 cells (\SI{2.54}{\centi\meter} per cell).
A total of 100 MGEP datasets were generated.
For each dataset, the control rods (layer H) were separated into the East and West half of the facility.
The control rods' location within each half was generated through a Latin hypercube sampler.
Therefore, for each MGEP dataset, we have a mapping $\mathbb{R}^2 \rightarrow \mathbb{R}^{4\times 90}$.

The MITR was constructed in 1956 and upgraded in 1974.
It is a light-water cooled reactor that operates at \SI{6}{\mega\watt} thermal.
However, it does not generate any electricity.
The primary utilization of the MITR is to irradiate experiments, either within the core or around the core periphery.
Although the total power of MITR is significantly lower than a commercial nuclear power plant, its in-core flux profile nearly matches \cite{mitrExp}.
Thus, the MITR enables development and testing of new materials, instruments and methods intended for use in commercial plants.
The MITR is a critical facility, which means that the number of neutrons created by fission is equal to those absorbed, \ie the neutron population is self-sustaining.
As the facility is critical, many active and passive safety features are incorporated into the MITR design, and each experiment undergoes a safety analysis before permitting in-core placement.

The cross-section of the MITR facility is presented in \cref{fig:systems}.
The core has a hexagonal cross-section, with rhomboidal fuel elements that are approximately \SI{0.6}{\meter} in height, and \SI{6.5}{\centi\meter} each side.
In total, there are 27 in-core locations, with 24 occupied by fuel elements and 3 by in-core experiment facilities.
In this work, we use an \pkg{MCNP5} \cite{team2003mcnp} model of the MITR, that has been experimentally validated \cite{sun2014validation}.
\pkg{MCNP5} is a closed-source code developed by the Los Alamos National Laboratory.
\pkg{MCNP5} also approximates \cref{eq:transport} through the MC method.
The MITR model is 3-D and consists of many components including the core, structural materials, graphite and heavy-water reflectors.
Modeling the material surrounding the core is necessary to account for nuclear transport into and out of the core.
The major control devices for the MITR are six shimblades located around the outer periphery of the core.
Similar to the MGEP control rods, the shimblades are strong neutron absorbers and their movement will perturb $\phi$.
A total of 151 MITR datasets were generated.
For each dataset, the shimblades' height was asymmetrically perturbed.
Using Latin hypercube sampling, each height had a center of \SI{24}{\centi\meter} and a uniform width of $\pm\SI{2}{\centi\meter}$.
The MCNP output is a 2-D tally for 22 fuel elements, discretized further into 16 axial nodes each. 
Therefore, for each MITR dataset, we have a mapping $\mathbb{R}^6 \rightarrow \mathbb{R}^{22\times 16}$.

\begin{figure}
\includegraphics[scale=0.262]{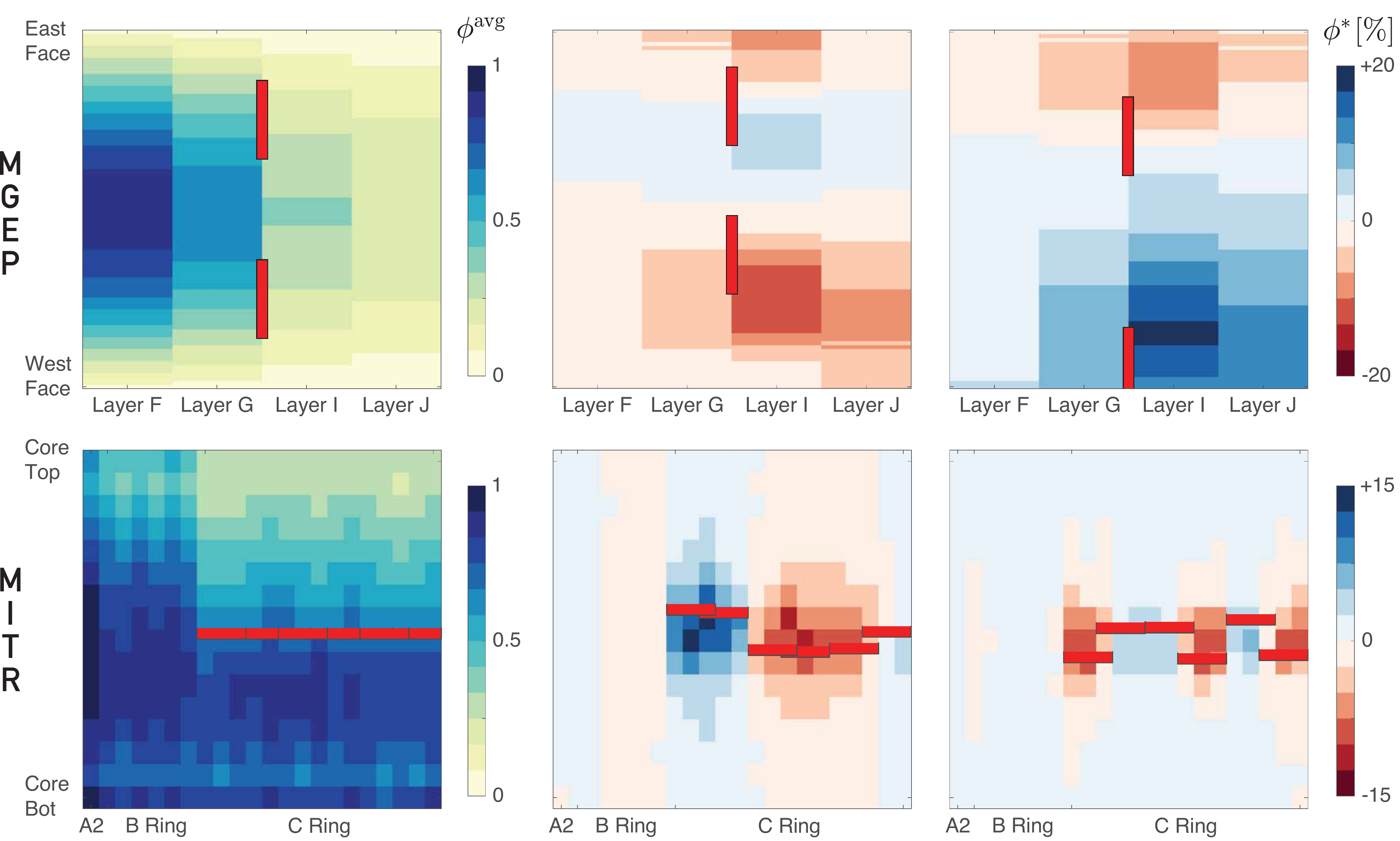}
\caption{A showcase of the neutron flux distribution and pertubation.
Left column: linearly scaled average flux distributions, $\phi^\mathrm{avg}$.
Red bars indicate approximate locations of the control devices.
For the MGEP, the lateral location of the control rod is indicated.
For the MITR, the height of the shimblade tip is indicated (blade inserted downwards, from core top).
Middle and right columns: examples of flux perturbation, $\phi^*$, due to movement of control devices from mean location.
}
\label{fig:flux}
\end{figure}

The movement of control devices in a fission system will perturb the spatial neutron flux distribution.
To illustrate and quantify these effects, the stationary neutron flux distribution and examples of perturbation are presented in \cref{fig:flux}.
The stationary distribution, $\phi^\mathrm{avg}$, is the average of all datasets generated, \ie the control devices are in their neutral positions (mean position used for Latin hypercube sampling).
The perturbed distribution, $\phi^*$, is the relative change due to the movement of control devices from their neutral positions.
For the MGEP, $\phi^\mathrm{avg}$ distribution follows our expectation of a subcritical system: $\phi$ is greatest at the location of the source and decays as the graphite lattice is traversed in both directions.
The axial position of the control rod significantly suppresses $\phi$ in layers I and J.
Moving the control rods towards the lateral center (between East and West faces) of the pile further suppresses $\phi$, and vice versa.
The local perturbation is greatest in layer I as neutrons are `streaming' towards the top of the pile, however we do observe lower order global perturbations throughout the system.
For the MITR, the $\phi^\mathrm{avg}$ distribution is more nuanced as we rely on sustenance of a chain reaction, rather than a point neutron source.
The shimblades are inserted from the top and around the C Ring elements.
This causes a depression in $\phi$ from the core top to the middle core.
Moving the shimblades up or down causes an increase or suppression on neighboring elements, and also perturbs the inner ring elements.
As discussed in \cref{appendix:A}, the solutions to simplified versions of \cref{eq:transport} for subcritical \vs critical systems vary qualitatively, and therefore we have an a priori expectation of different perturbation dynamics.

\FloatBarrier
\subsection{Objective}

This work contrasts several ML algorithms to generate empirical models for multidimensional regression of neutron transport in fission systems.
ANN, GBR, GPR, and SVR is explored.
MC solutions of neutron transport in the MGEP and the MITR are used as datasets.
Each model will provide a function,
\begin{equation}
f:\mathbb{R}^c \rightarrow \mathbb{R}^{e\times n}~,
\label{eq:map}
\end{equation}  
where $c$ is the number of control devices, and $e\times n$ is the shape of the neutron transport tally, \ie the discrete MC approximation of \cref{eq:transport}.
To eliminate the subjectivity of hyperparameter selection, we employ a systematic meta-learning procedure.
In \cref{sec:methods}, the ML algorithms and metrics used to quantify performance are discussed.
In \cref{sec:metalearning}, findings from hyperparameter optimization are presented.
In \cref{sec:results}, results, including physical implications, are detailed.
Concluding remarks from this work are presented in \cref{sec:conclusions}.

\section{Methods}\label{sec:methods}

In pursuit of maximizing accuracy, precision and capability to generalize, we contrast several ML regression algorithms in this work.
For each algorithm, we leverage existing libraries.
For ANN, we use \pkg{Tensorflow}/\pkg{Keras} \cite{tensorflow2015-whitepaper,chollet2015keras}.
For GBR and SVR, we use \pkg{scikit-learn} \cite{scikit-learn2011}.
For GPR, we use \pkg{GPflow} \cite{GPflow2017,GPflow2020multioutput}.
Each package has expended significant effort in abstracting the process of generating models.
However, the issues of hyperparameter tuning and the input/output pipeline requires additional layers.
To handle these tasks, specifically for our context of nuclear transport regression, we have developed an open-source python package \pkg{NPSN} (\href{https://github.com/a-jd/npsn}{github.com/a-jd/npsn}).

\subsection{ML Algorithms}\label{subsec:models}

The first regression algorithm considered was the ANN.
ANNs have been the centerpiece of recent ML development, advancing the state-of-the-art in computer vision, speech recognition, natural language processing, \etc \citep[Ch. 12]{goodfellow2016deep}.
Our previous work \citep{dave2020deep} indicated ANNs had promising performance in multidimensional regression of the MITR.
The particular classification of ANNs utilized in this work are known as feedforward neural networks, where information flows from the input to the output layer (with no backward/feedback connections).
The regression function is built through connecting several functions in a chain,
\begin{equation}
f(\mathbf{x}) = f_3(f_2(f_1(\mathbf{x})))~,
\label{eq:ann}
\end{equation}
where $f_1$ is the input layer, $f_2$ is the intermediate layer, and $f_3$ is the output layer.
For each layer, the transformation takes the form,
\begin{equation}
f(\mathbf{x}) = g(\mathbf{w}^\intercal\mathbf{x}+\mathbf{c})~
\end{equation}
where $\mathbf{w}$ are weights for a linear transformation, $\mathbf{c}$ is the bias vector, and $g(\cdot)$ is known as the activation function.
Appropriately choosing an activation function has a significant impact on the capability to address non-linear problems.
Additionally, there are other choices available to the user, such as the number of layers and the shape of the intermediate layers itself.

We implemented GBR, an ensemble method.
Ensemble methods combine multiple estimators with an expectation of improved generalization over single estimators.
Within ensemble methods, boosting \citep{freund1997decision} builds multiple estimators sequentially.
The GBR estimator is defined as,
\begin{equation}
f_M(\mathbf{x}) = \sum_m^M h_m(\mathbf{x})~,
\label{eq:boosting}
\end{equation}
a sum of $M$ estimators, $h_m$.
Each estimator is determined sequentially,
\begin{equation}
f_m(\mathbf{x}) = f_{m-1}(\mathbf{x}) + h_m(\mathbf{x})~,
\end{equation}
where $h_m$ is determined greedily by minimizing the gradient of the loss function, $l(y_i, f(x_i))$, 
\begin{equation}
h_m \approx \argmin_h \sum_{i=1}^n h(x_i)\left[\frac{\partial l\left(y_i,f_{m-1}(x_i)\right)}{\partial f_{m-1}(x_i)}\right]~.
\end{equation}
Therefore, each boosting stage is considered as a gradient descent in the loss function's space.
Other ensemble methods such as bagging were not explored in this work.
Additionally, Gradient Boosting was chosen over other tree-based methods such as random forrests (due to benefits in performance reported in other regression problems \citep{ogutu2011comparison,sahin2020assessing}).

Next, we explored the use of SVR.
Support Vector Machines were proposed by \citet{cortes1995support} for the classification problem.
However, significant work has extended their application towards regression.
In particular, we implement the \nusvr model proposed by \citet{scholkopf2000new}.
The SVR model can be written in the form,
\begin{equation}
f(\mathbf{x}) = b + \sum_i \alpha_i k\left(\mathbf{x}, \mathbf{x}_i\right)~,
\label{eq:svr}
\end{equation}
where $b$ is an offset, $\mathbf{x}$ is the input data, $\mathbf{x}_i$ is a training sample, and $\alpha_i$ is a trained coefficient vector.
The major concept introduced was the so-called \textit{kernel trick} where $k(\mathbf{x},\mathbf{x_i})=g(\mathbf{x})^\intercal g(\mathbf{x}_i)$ allowed arbitrary linear and nonlinear kernels, $g(\mathbf{x})$, to be utilized in SVR.
The kernel trick allows the relationship between $b, \mathbf{\alpha}, g(\mathbf{x})$ and $f(\mathbf{x})$ to remain linear, while that between $\mathbf{x}$ and $g(\mathbf{x})$ may be nonlinear.
Thus, the SVR algorithm is able to learn models that are nonlinear functions of $\mathbf{x}$, while utilizing convex optimization techniques guaranteed to converge efficiently \citep[Ch. 5.7.2]{goodfellow2016deep}.
The \nusvr model improves over the traditional SVR algorithm by allowing users more control over the proportion of support vectors retained (\vs total number of samples).

Lastly, we also explored GPR, adding a stochastic approach to our regression problem.
Gaussian processes \cite{goertler2019a} are a non-parametric approach to regression.
That is, unlike the algorithms discussed above, where parameters are tuned to produce a regression function, in Gaussian processes \textit{distributions} of parameters are found to produce \textit{distributions} of possible functions.
To achieve this, GPR follows the Bayesian method of beginning with a prior distribution, and updating it through data exposure to produce a posterior.
A Gaussian process is defined by the property that function values are Gaussian distributed,
\begin{equation}
p(f(\mathbf{x})) = \mathcal{N}(f(\mathbf{x}), \boldsymbol{\mu}, \mathbf{K})~,
\label{eq:gpr}
\end{equation}
where $\boldsymbol{\mu}$ represents the mean values, and $\mathbf{K}$ represents the covariance matrix.
Normalizing the input distribution leads to $\boldsymbol{\mu}\approx\mathbf{0}$.
The covariance matrix is determined by the kernel which evaluates the similarity (influence) between two points.
Ultimately, the kernel describes the shape of the posterior distribution and thus, the characteristics of $f(\mathbf{x})$.
For our particular problem of multiple-output Gaussian process, we use the Sparse Variational Gaussian Process \citep{GPflow2020multioutput}.

It is important to note that GBR and SVR algorithms output scalar values natively.
To output vectors, an aggregate of \cref{eq:boosting,eq:svr} is needed (in accordance with \cref{eq:map}, $e\cdot n$ instances).
To achieve this, the \pkg{MultiOutputRegressor} in \pkg{scikit-learn} was utilized to wrap multiple GBR and SVR models.
At the least, computational overhead for GBR and SVR algorithms will scale $\propto e\cdot n$.
Whereas the ANN and GPR algorithms output vector values natively.

\subsection{Package Development}

\begin{figure}
\includegraphics[scale=0.75]{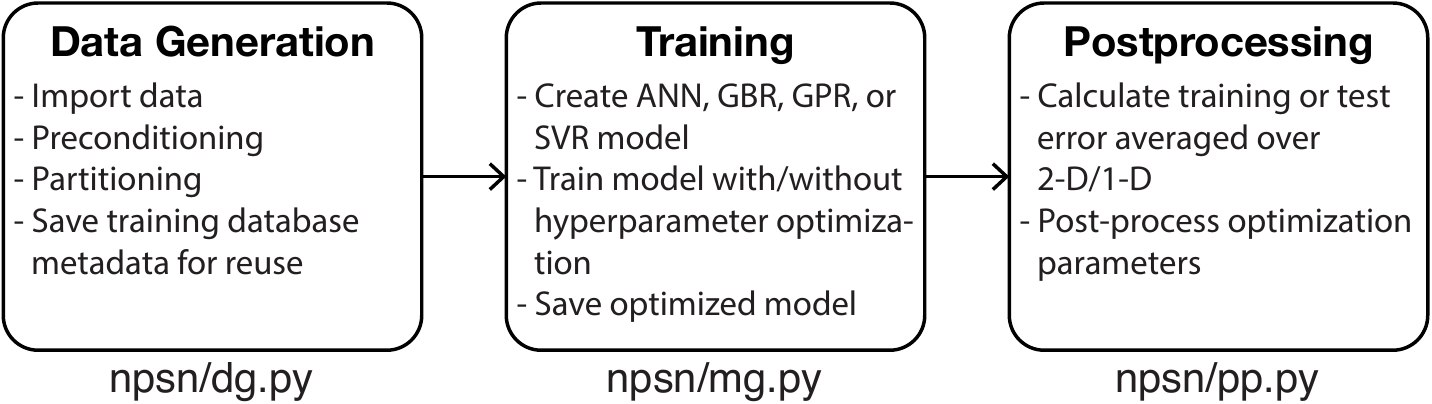}
\caption{Organization of major components of the \href{https://github.com/a-jd/npsn}{\pkg{NPSN}} package.}
\label{fig:npsn}
\end{figure}

To guide the development of \pkg{NPSN}, we established several design goals.
Primarily, we want to reduce friction in going from a dataset that follows our mapping ($\mathbb{R}^c \rightarrow \mathbb{R}^{e\times n}$), to obtaining an optimized ML model.
To address this goal, we made a significant effort in abstracting the package interface (example input provided in \cref{appendix:B}).
Second, we want \pkg{NPSN} to serve as a template for other developers, reducing the effort necessary in developing custom pipelines.
To address this goal, we use the paradigm of class inheritance to allow easy integration of other user-defined ML algorithms.
Our final goal is that \pkg{NPSN} is continuously optimized and remains compatible with major revisions of dependencies.
In developing \pkg{NPSN} as an open-source package, we hope to receive feedback and constructive critique from the community.
A visualization of the package layout is presented in \cref{fig:npsn}.

\subsection{Training \& Performance Metrics}

In order to consistently contrast performance of all models explored, a common set of training and performance metrics is defined.
During training and meta-learning, a scalar value to guide the optimization process is required.
The training error for a combination of hyperparameters, $\theta$, is defined by the mean squared error,
\begin{equation}
\gamma_\theta(\hat{y},y) = \frac{1}{n}\sum_{ijk}(\hat{y}_{ijk}-y_{ijk})^2~,
\label{eq:train}
\end{equation}
where $\hat{y}$ is the ML model predicted $\phi$, and $y$ is the \pkg{OpenMC} or \pkg{MCNP5} solution.
The subscript $i$ represents the element node, $j$ the axial node, and $k$ is a permutation index of the control device locations.
The denominator $n$ is calculated by $n=\max(i)\max(j)n_{\mathrm{test}}$.
For the MGEP system, $\max(i)=90$ and $\max(j)=4$.
For the MITR system, $\max(i)=22$ and $\max(j)=16$.
The total number of control permutations explored was 155 for the MGEP, and 100 for the MITR.
The training \vs test sets had an 80-20 split.
Thus, for the MGEP and MITR, $n_\mathrm{test}$ was 31 and 20, respectively.
The meta-learning process evaluates multiple combinations of $\theta$ to arrive at an optimized model.
Through evaluating \cref{eq:train} on \textit{test} datasets only, the optimizion process is expected to produce a model that generalizes well, \ie perform well on ``unseen'' data. 

The performance of the optimized model is quantified by two metrics.
First, the accuracy of the trained model is defined by the mean absolute percentage error,
\begin{equation}
\varepsilon_{ij}(\hat{y},y)  = \frac{1}{K}\sum_{k=1}^K\left|\frac{\hat{y}_{ijk}-y_{ijk}}{y_{ijk}}\right|~,
\label{eq:acc}
\end{equation}
retaining above definitions of $\hat{y},~y$, and $K$ is the total number of training or test permutations.
If a node-averaged distribution, $\varepsilon_i$, an element-averaged distribution, $\varepsilon_j$, or a domain-averaged distribution $\varepsilon$, is sought, additional averaging is performed.
In addition to accuracy, the precision of the model is quantified.
The precision is an important consideration as large variances in model outcome could lead to an unstable controller.
Precision is quantified by the standard deviation of the error,
\begin{equation}
\sigma_{ij}(\hat{y},y)= \sqrt{\frac{1}{K}\sum_{k=1}^K\left(\varepsilon_{ijk}-\varepsilon_{ij}\right)^2}~,
\label{eq:std}
\end{equation}
where $\varepsilon_{ijk}$ is the error before averaging over $k$.
Likewise, additional averaging could produce $\sigma_i$, $\sigma_j$ or $\sigma$.
In \pkg{npsn/pp.py} (\cref{fig:npsn}) we implement \cref{eq:acc,eq:std}.
Sharing the post-processing module enforces an unbiased comparison of all models.
By choosing the subset of $k$ that \cref{eq:acc,eq:std} are evaluated on, one can determine the training, $\varepsilon_{ij}^\mathrm{Train}$, or test error, $\varepsilon_{ij}^\mathrm{Test}$.
The generalization error is determined by,
\begin{equation}
\varepsilon_{ij}^\Delta=\varepsilon_{ij}^\mathrm{Test}-\varepsilon_{ij}^\mathrm{Train}~.
\label{eq:gen}
\end{equation}

\cleardoublepage
\section{Meta-learning}\label{sec:metalearning}

Parametric ML algorithms have a large degree of freedom in configuration settings.
These settings are user-defined and do not change after each training iteration.
The settings are known as hyperparameters, and their choice makes a significant impact on performance.
The optimization of hyperparameters is known as meta-learning.
\pkg{NPSN} utilizes the optimization library \pkg{HyperOpt} \cite{Bergstra2013} which optimizes over awkward search spaces.
For hyperparameters, these awkward spaces include real-valued, strings and boolean dimensions.
The Tree of Parzen optimizer \cite{bergstra2011algorithms} is utilized  which has shown superior performance over random optimizers for computer vision problems.

All the models considered in this work have several hyperparameters with default settings, but require further tuning to arrive at an optimized model.
The subset of settings considered in this study, for each model, is tabulated in \cref{table:hypss}.
For ANN, GBR and SVR, over 5 parameters are considered for optimization.
For GPR, which is a non-parameteric model, the parameters are considerably lower.
Non-parametric does not imply that parameters do not exist, but rather that the approach finds distributions of parameters, which have infinitely many points.
From the data types column, it is recognized that the search space is indeed `awkward' as a mix of real-valued data and strings (representing selections of kernels, \etc) are present.
Detailed description of each parameter's functionality and available options is deferred to the online manual of each library referenced in the \cref{sec:methods} introduction.

\begin{table}
\caption{Hyperparameter search space and optimal settings for each system and algorithm combination.}
\label{table:hypss}
\resizebox{\linewidth}{!}{
\begin{tabular}{llllcc}
\hline
& & & & \multicolumn{2}{c}{Optimal Settings} \T\\
 & Parameter & Type & Range & MGEP & MITR \B\\
\hline
ANN & IDL Number of Layers & Integer & 0-5 & 5 & 0  \T\\
& IDL Shape of Layers & Integer & 1-4 times input shape & 4 & - \\
& IDL Activation Function & String & tanh, SoftPlus, SoftSign, Sigmoid, ReLU, ELU & \multicolumn{2}{c}{ReLU} \\
& Optimizer Type & String & SGR, RMSprop, adam & \multicolumn{2}{c}{adam} \\
& Network Loss Function & String & MSLE, MAPE, MSE, logcosh & logcosh & MSE \\
& Batch Size & Integer & [4,8,16,32] & 8 & 4 \B\\
\hline
GBR & Loss Function & String & Huber, LAD, LS & LAD & Huber \T\\
& Learning Rate & Float & 0.05-0.4 & 0.08 &  0.10 \\
& Boosting Stages & Integer & 100-400 & 312 & 368 \\
& Split Criterion & String & MAE, MSE, FMSE & \multicolumn{2}{c}{FMSE} \\
& Maximum Depth & Integer & 2-10 & 7 & 2 \\
& Maximum Features & String & `auto', `sqrt', `log2' & \multicolumn{2}{c}{'auto'} \B\\
\hline
GPR & Kernel & String & 
\mtl{
Linear, Exponential, Matern52, \\
Linear+Exponential, Linear$\times$Exponential,\\
D-Linear+Exponential, D-Linear$\times$Exponential,\\
Linear+Matern52, Linear$\times$Matern52,\\
D-Linear+Matern52, D-Linear$\times$Matern52,\\
Exponential+Matern52, Exponential$\times$Matern52,\\
D-Exponential+Matern52, D-Exponential$\times$Matern52\\
} & Exponential & \mtc{Exponential+\\Matern52}
\T\\
& Inducing Points & Integer & 21, 45, 75, 101 & 101 & 45 \B\\
\hline
SVR & Kernel Type & String & Sigmoid, RBF, Polynomial, Linear & \multicolumn{2}{c}{RBF} \T\\
& $\nu$ (Vector Retention) & Float & $10^{-5}$-1.0 & 0.52 & 0.40 \\
& C (Regularization Parameter) & Float & 0.5-10.0 & 9.82 & 1.91 \\
& $\gamma$ (Kernel Coefficient) & String & `auto' or `scale' & `scale' & `auto' \\
& Polynomial Degree & Integer & 2-5 (only valid for polynomial kernel) & \multicolumn{2}{c}{-} \B\\
\hline
\end{tabular}
}
\end{table}

\begin{landscape}
\begin{figure}
\captionsetup{width=6.5in}
\includegraphics[width=\linewidth]{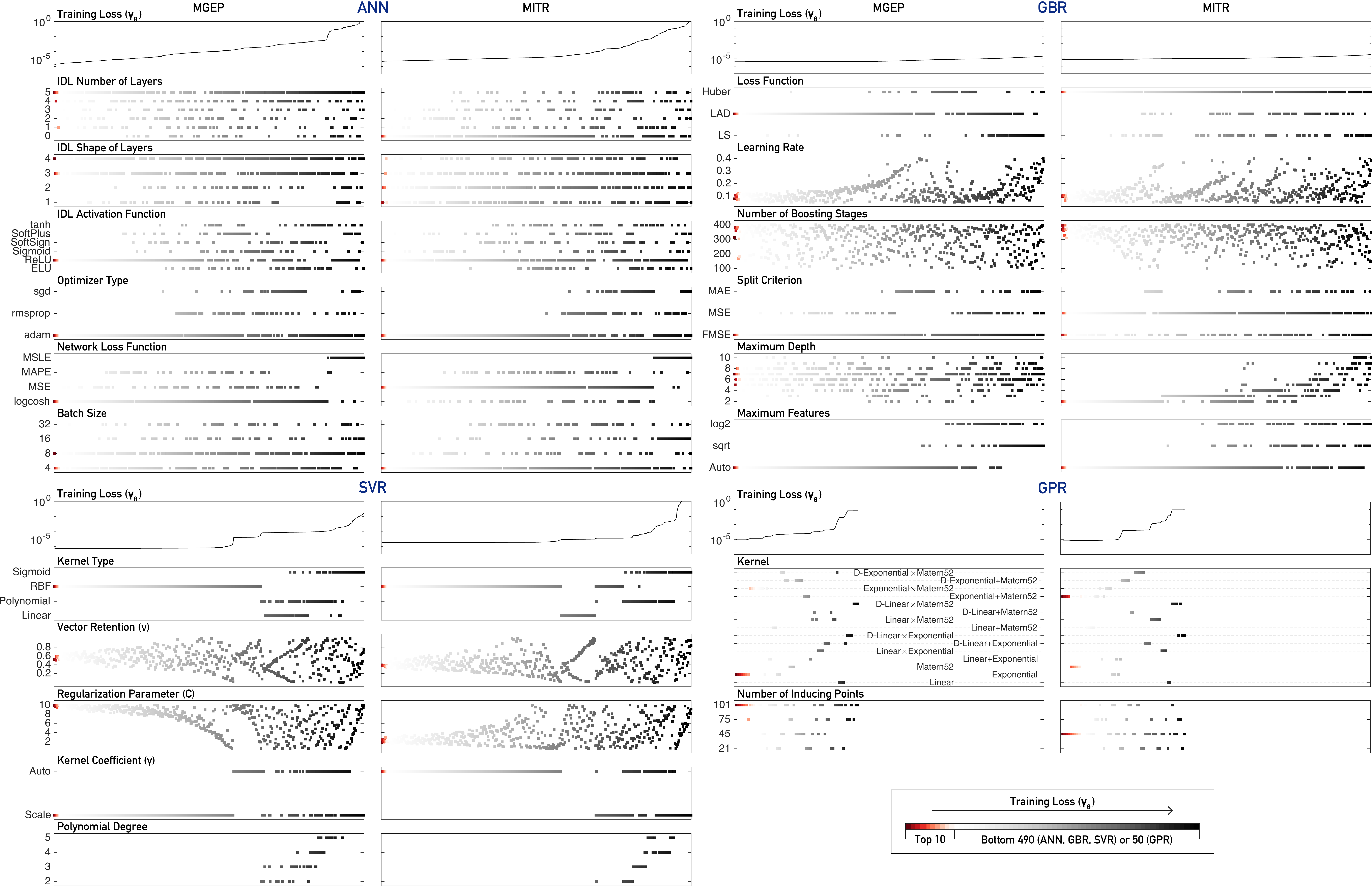}
\caption{An overview of the hyperparameter optimization results for each algorithm and system.
Combinations are ranked in ascending training loss, $\gamma_\theta$.
The top 10 configurations are highlighted in a red gradient, and the remaining in a gray scale gradient, as illustrated on the bottom right hand corner.}
\label{fig:opt-comp}
\end{figure}
\end{landscape}

The optimization process reveals several interesting findings.
An overview of the optimization outcome is compiled in \cref{fig:opt-comp}.
The outcome for each model will be discussed separately, followed by a discussion of intersecting trends across the models.

The optimization of ANN architecture involved setting the optimizer type, network loss function, batch size, and configuration of the intermediate dense layer (IDL).
The IDL is the intermediate function(s), $f_2$, in \cref{eq:ann}.
Manipulations include the depth (number of IDL layers), shape, and its activation function.
First, the shared outcomes between MGEP and MITR systems is discussed.
Both systems benefit from lower batch sizes, the logcosh network loss function, the adam optimizer, and the ReLU activation function.
Whereas the settings differ in terms of the shape and depth of the IDL, which is interpreted as the capacity of the ANN.
The MGEP model benefits from a larger capacity network, whereas for MITR, the opposite is true.
An interesting outcome is that this trend continues for the remaining ML algorithms.
Lastly, another observation is that the training loss, $\gamma_\theta$, is almost asymptotic for the MITR.
Whereas the MGEP model may benefit from further increases in the IDL layer depth.

The optimization of the GBR model involved setting the loss function, learning rate, number of boosting stages ($M$ in \cref{eq:boosting}), the split criterion, maximum tree depth, and maximum features.
Both systems benefit from a learning rate $\approx 0.1$, a high number of boosting stages, and the same method of determining maximum features.
However, there are differences in optimal choices for the split criterion and loss function.
Again, the MGEP system benefits from a greater ``depth'' (\ie the depth of regression trees used in each boosting stage).

The optimization of the SVR model involved setting the kernel type, vector retention parameter, regularization parameter, kernel coefficient, and when applicable, the polynomial degree.
The kernel type manipulates $k$ in \cref{eq:svr}.
Both systems benefit from the Radial Basis Function (RBF) kernel.
In fact, the greatest improvement in $\gamma_\theta$, is from selection of RBF as the kernel.
The systems differ in optimal choices of all other parameters.
However, the other parameters have a minor impact on improvement of $\gamma_\theta$.

The optimization of the GPR model involved setting the kernel and the number of inducing points.
The non-parametric designation is apparent.
The number of inducing points is the number of points on which the kernel is defined and then trained.
The number of optimal inducing points for the MITR is lower than that for the MGEP -- in fact, it is less than half.
This was analogously noted for the ANN and GBR models, where network capacity and tree depth, respectively, had similar outcomes.
The kernel function used in GPR determines the covariance matrix, $\mathbf{K}$ in \cref{eq:gpr}.
The kernel controls the underlying multivariate distribution that the fitted model can adopt, and thus, dictates the function characteristics. 
There are several kernel options explored.
When a kernel is simply `Linear' or `Exponential', the same kernel is applied to both the $e$ and $n$ dimensions, in \cref{eq:map}.
When two kernels are stated, a summation or product of kernels is applied to both input dimensions, \eg the `Linear+Exponential' kernel applies a sum of linear and exponential kernels; `Linear$\times$Exponential' kernel applies a product of the linear and exponential kernels.
When two kernels are prefixed with a `D-', we separate the dimensions on which the kernels are activated, \eg the `D-Linear+Exponential' kernel applies a sum of the linear and exponential kernels where the linear kernel is activated by the $e$ dimension and the exponential kernel is activated by the $n$ dimension.
The MGEP model benefits significantly from the `Exponential' kernel.
Whereas the MITR model is optimal with either the `Exponential+Matern52' or `Matern52' kernel.
The Mat\'ern \citep[Ch. 4.2]{williams1996gaussian} covariance function is based on Bessel functions.
Therefore, the GPR optimization outcome aligns with a priori knowledge that the underlying characteristics of neutron transport between both systems differ.

The meta-learning process provides several important findings:
\begin{itemize}
\item 
A priori, the solution of \cref{eq:transport} between a subcritical (MGEP) and critical (MITR) system is expected to differ qualitatively, \eg \cref{eq:subsol} \vs \cref{eq:supsol}.
This manifests in differing optimal hyperparameter sets between both systems, for each ML model considered.
Thus, if we \textit{know} that systems we are modeling have qualitatively different solutions, we need to conduct separate meta-learning processes.
To confirm that criticality is the major driver for differing sets, further work is warranted to compare subcritical/critical systems with different designs (materials choices, energy spectra, and geometry).
\item
A common trend across the ANN, GBR and GPR models was found regarding the capacity of the model.
The optimal MGEP model required a greater capacity than the MITR equivalent.
This is an interesting finding that may be related to the respective mapping.
For the MGEP, each model maps $\mathbb{R}^2 \rightarrow \mathbb{R}^{4\times 90}$, and for the MITR $\mathbb{R}^6 \rightarrow \mathbb{R}^{22\times 16}$.
The output space for both systems is roughly equivalent, whereas the input space for the MGEP is a third of the MITR's.
Heuristically, it can be concluded that a larger model capacity is needed for a larger transition in dimensional mapping.
\item
Another finding concerns the range of training loss observed.
In \cref{fig:opt-comp}, the hyperparameter sets, $\theta$, are ranked in ascending training loss, $\gamma_\theta$.
For ANN, SVR, and GPR, the difference in $\gamma_\theta$ between the upper and lower decile $\theta$, is several orders of magnitude.
Observing the $\theta$ across the poorly performing deciles shows that there is no coherent pattern in guaranteeing high $\gamma_\theta$ (\ie a poorly performing model).
In other words, hand-tuning $\theta$ will result in erratic outcomes and may often discourage users from certain ML algorithms.
Therefore, a systematic meta-learning process is necessary in obtaining optimal model performance.
However, the GBR algorithm is less sensitive to the hyperparameters chosen, but have other drawbacks which will be discussed in the next section.
\end{itemize}

\section{Performance of Optimized Models}\label{sec:results}

The meta-learning procedure culminates in optimized ML models for each system.
The optimal settings for each combination are tabulated in \cref{table:hypss}, and are used to produce all results discussed in this section.
The metrics used to quantify performance are the accuracy, precision, and generalization error, determined by \cref{eq:acc,eq:std,eq:gen}, respectively.
First, a summary of the outcome is presented.
In \cref{subsec:spatial}, the spatial performance distribution and its physical significance is discussed.
In \cref{subsec:data}, the incremental improvement of performance, with respect to dataset size is discussed.

After the systematic meta-learning process, it is assumed that improvements to model performance are exhausted and an optimal model is obtained.
In other words, we assume further tuning of the hyperparameters will not result in any improvements, only fundamental changes to the underlying ML algorithm may.
A summary of the performance for each ML model is tabulated in \cref{table:perf}.
The proceeding statements are valid for both MGEP and MITR systems.
The SVR model has the greatest accuracy and precision.
The GPR model has the lowest generalization error.
The GBR model has the largest generalization error.
The ANN model has the lowest execution time, by an order of magnitude.
The GPR model has the greatest execution time, by several orders of magnitude.
In general, it is found that all ML algorithms provide more accurate and precise surrogates of the MITR than the MGEP.
This result may be attributed to the greater magnitude of perturbation (\eg \cref{fig:flux}) in the MGEP.

\begin{table}
\caption{A comparison of domain-averaged performance metrics for both systems and all models.
The test error, generalization error, precision and average time required for a single evaluation, $T$, are presented.}
\label{table:perf}
\begin{tabular}{c@{\hskip 24pt}cccc@{\hskip 24pt}cccc}
\hline
& \multicolumn{4}{c}{MGEP} & \multicolumn{4}{c}{MITR} \T\\
& $\varepsilon^\mathrm{Test}~[\%]$ & $\varepsilon^\Delta~[\%]$ & $\sigma^\mathrm{Test}~[\%]$ & $T$ [\SI{}{\milli\second}] 
& $\varepsilon^\mathrm{Test}~[\%]$ & $\varepsilon^\Delta~[\%]$ & $\sigma^\mathrm{Test}~[\%]$ & $T$ [\SI{}{\milli\second}] \B\\
\hline
ANN & 0.371 & 0.056 & 0.231 & 0.093  & 0.215 & 0.037 & 0.150 & 0.023 \T\\
GBR & 0.466 & 0.461 & 0.396 & 3.62   & 0.283 & 0.259 & 0.213 & 1.49 \\
GPR & 1.075 & 0.004 & 0.730 & 1.48   & 0.237 & 0.026 & 0.168 & 0.723 \\
SVR & 0.170 & 0.045 & 0.131 & 0.677  & 0.172 & 0.048 & 0.128 & 0.675 \B\\
\hline
\end{tabular}
\end{table}
 
The results provide guidance for selection of ML algorithms for multidimensional regression of fission systems.
First, although the GBR algorithm does not require exhaustive hyperparameter optimization, it suffers from large generalization error and is not recommended.
Next, although the GPR algorithm generalizes the best, it has poor accuracy and requires the greatest computational overhead by several orders of magnitude.
We are left with two options: ANN and SVR.
The choice between both options would need to weigh the accuracy/precision advantage of the SVR over the computational efficiency of ANNs.\footnote{To remove any advantage due to GPU acceleration, all models were evaluated using the CPU only.
An Intel 9900K CPU was used to evaluate all models.
Another caveat is that alternatives libraries may be more efficient than those adopted in this work.
Thus, an exhaustive evaluation of all libraries and acceleration with GPUs is recommended for further optimization.}
Both ANN and SVR require hyperparameter optimization.

\subsection{Spatial Distribution}\label{subsec:spatial}

\begin{figure}
\resizebox*{!}{0.925\textheight}{
\begin{tabular}{c}
\includegraphics[width=\linewidth]{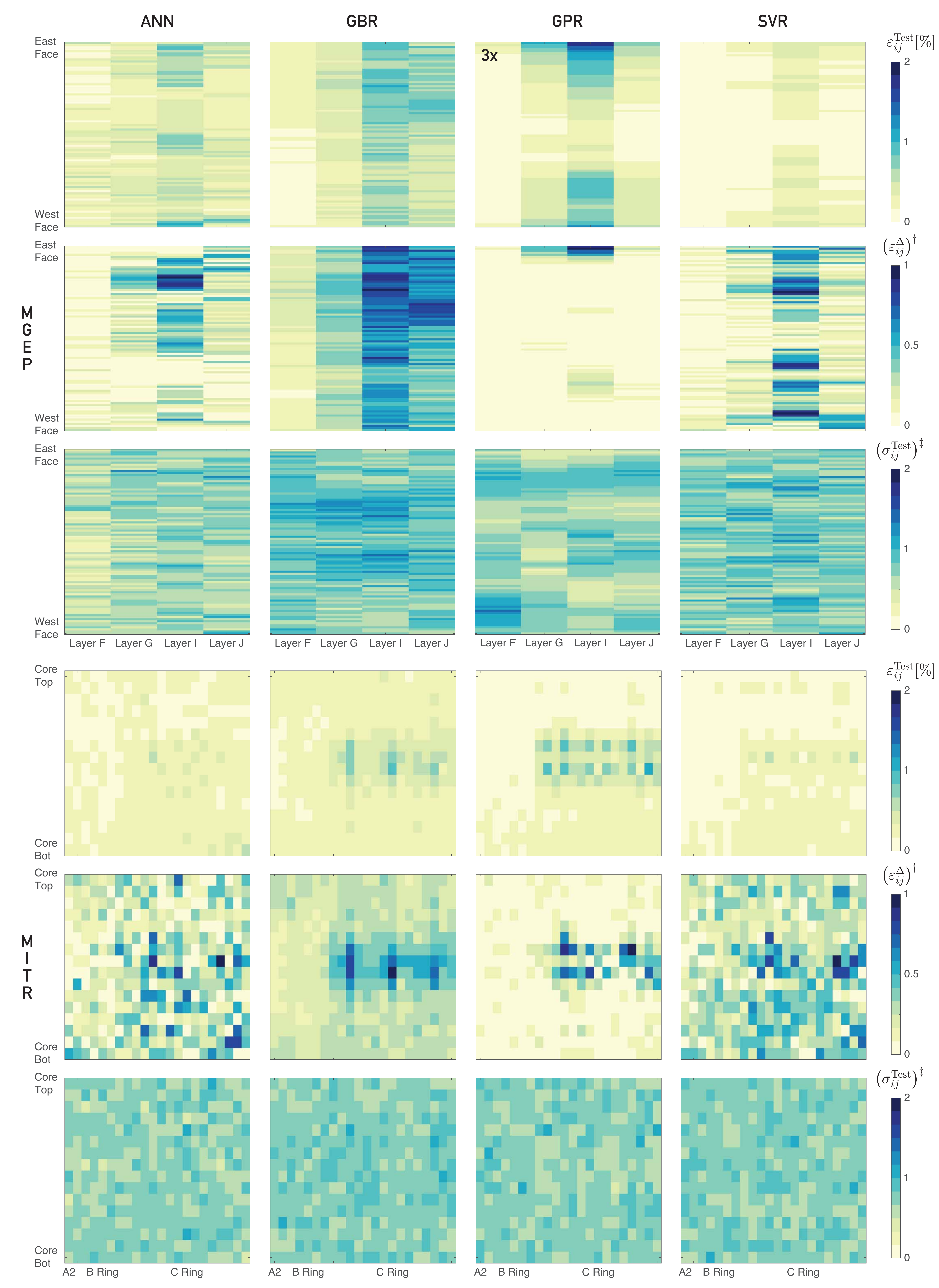}
\end{tabular}
}
\caption{Spatial distribution of test accuracy, \cref{eq:acc}, generalization error, \cref{eq:gendagger}, and test precision, \cref{eq:stddagger}.
The $\varepsilon_{ij}^\mathrm{Test}$ range of the MGEP GPR model is triple of the values represented by its colorbar (designated by `3x').
}
\label{fig:error}
\end{figure}

The spatial characterization of performance provides further insight in assessing ML surrogates, in the context of fission systems.
A compilation of the accuracy, generalization error and precision is presented in \cref{fig:error}.
The generalization error and precision have been transformed.
The generalization error is transformed such that,
\begin{equation}
\left(\varepsilon_{ij}^\Delta\right)^\dagger=\varepsilon_{ij}^\Delta/\max(\varepsilon_{ij}^\Delta)~,
\label{eq:gendagger}
\end{equation}
where $\varepsilon_{ij}^\Delta$ is determined by \cref{eq:gen}.
The precision is transformed by a Hadamard division,
\begin{equation}
\left(\sigma_{ij}^\mathrm{Test}\right)^\ddagger=\sigma_{ij}^\mathrm{Test}/\varepsilon_{ij}^\mathrm{Test}~,
\label{eq:stddagger}
\end{equation}
where $\varepsilon_{ij}^\mathrm{Test}$ is determined by \cref{eq:acc}, and $\sigma_{ij}^\mathrm{Test}$ by \cref{eq:std}.

The spatial distribution of the accuracy on test data, $\varepsilon_{ij}^\mathrm{Test}$, is discussed first.
At first glance, it is apparent that there are coherent patterns representing spatial regions for both systems where all ML models struggle.
For the MGEP, spatial regions of poor performance tend to occur in Layer I, the layer that is directly above the control rods (\cref{fig:systems}).
For the MITR, spatial regions of poor performance occurs in the C-ring, the ring that is adjacent to the shimblades.
Additionally, the spatial region corresponds with the center of the core (\ie between `Core Top' and `Core Bot').
We have previously highlighted that the MGEP and MITR are different qualitatively due to their criticality differences.

\FloatBarrier
In a subcritical system, the neutron source location is the primary driver of $\phi(\mathbf{x})$.
In the MGEP, the neutron source is located below layer A; neutrons transport upwards towards the top of the pile.
As neutrons pass between layer G and I, they may encounter the control rods and get absorbed, depressing $\phi$ locally.
This local depression impacts ``downstream'' layers I and J.
An analogy here would be the solar eclipse, where light is obstructed by the moon.
We note that all ML models perform well in the two layers ``upstream`` of the control rods (F and G), and perform worst in the I layer followed by J.
The SVR model performs the best, showing moderate error of $\approx0.5\%$ in Layer I.
The GPR model performs the worst, showing up to $6\%$ error in Layer I, especially towards the East and West faces.

In a critical system, the neutron population has both localized and global effects as the reaction is self-sustaining.
Therefore, manipulating the MITR shimblades will impact the local and global $\phi(\mathbf{x})$ distribution.
The MITR shimblade locations (\ie the input for \cref{eq:map}) were sampled such that the mean physical tip of the blades corresponded to the center of the MITR core.
Therefore, we note that $\varepsilon_{ij}^\mathrm{Test}$ is greatest in locations where we expect local perturbation to be greatest.
The non-linear global effects are less apparent in the results.
We note that all ML models, again, perform well in locations where perturbation is expected to be low.
The SVR and ANN models perform the best, with $<1\%$ error in the high perturbation region.
The GBR and GPR models perform poorly, with $>1\%$ error.

The spatial distribution of the generalization error, $\varepsilon_{ij}^\Delta$, is discussed next.
For the MGEP, in descending order, the SVR, ANN and GBR struggle with generalizing neutron transport in Layers I and J.
The GPR model generalizes well and has most difficulty in the East Face region (this anomaly may be explained by a lack of samples towards the East face, an extremum for the Latin hypercube).
For the MITR, the GBR and GPR models retain similar spatial distributions as $\varepsilon_{ij}^\mathrm{Test}$.
Whereas the ANN and SVR models have a rather uniform spatial distribution.
The latter observation, combined with the low domain-averaged error ($\varepsilon^\Delta$ in \cref{table:perf}), suggests that the ANN and SVR models have accurately captured the dynamics of \cref{eq:map} locally and globally.

Finally, we discuss the precision on test data, $\sigma_{ij}^\mathrm{Test}$.
For the MITR, we observe no coherent spatial pattern, and remarkably similar normalized values, across all ML models.
For the MGEP, the outcome is somewhat similar, with the ANN model exhibiting better precision in the layers closer to the neutron source.
These results indicate that the precision, as defined in \cref{eq:std,eq:stddagger}, does not offer additional spatial information.
The explicit outputs of the ANN, GBR and SVR models, \cref{eq:ann,eq:boosting,eq:svr}, do not specify variance with respect to the output.
However, the explicit output of the GPR model, \cref{eq:gpr}, does and is discussed next.

\begin{figure}
\includegraphics[scale=0.262]{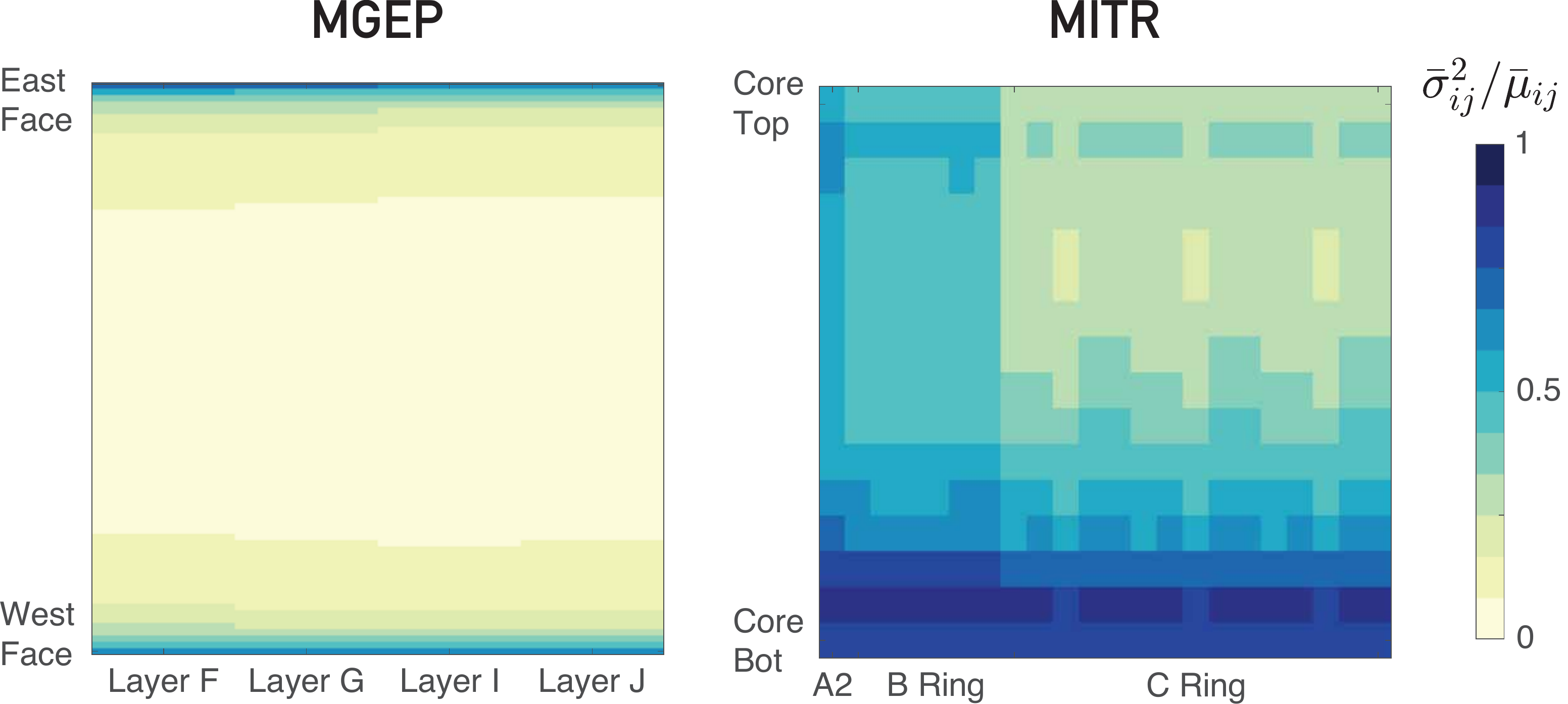}
\caption{Inference of variance by the GPR model. The scaled variance, $\bar{\sigma}_{ij}^2$, normalized by the scaled mean, $\bar{\mu}_{ij}$, for the MGEP and MITR.
The values have been averaged over all test data sets.}
\label{fig:gpr}
\end{figure}

The GPR model's output is a multivariate posterior probability distribution.
Therefore, we can sample both the mean and the variance after training.
In other models, regression would provide estimates of $f(\mathbf{x})\sim\phi(\mathbf{x})$, and no additional information on the confidence in the estimates.
Indirectly, we were inferring precision using \cref{eq:std}, which only quantified the stability of the model to accurately infer $f(\mathbf{x})$ across test data.
In \cref{fig:gpr}, the scaled variance for both systems is presented.
For the MGEP, the variance is very low in the middle of the system, and high at the East and West boundaries.
For the MITR, the variance is low in the C-ring, and high towards the inner elements.
Physically, the variance is low in spatial regions that are perturbed by control rod movement.
In other words, the GPR model has greater confidence in spatial regions that are perturbed by the model input.
The inference of variance, along with the mean, is particularly useful in safety analyses of critical fission systems where quantification of confidence intervals is necessary.
To reduce variance in regions of low perturbation, the prior variance requires adjustment.
In future work, the prior uncertainty in the regions of low perturbation will be determined through experimental data.

\FloatBarrier
\subsection{Dataset Requirements}\label{subsec:data}

The database requirements for training ML algorithms are not quantitative and rely on heuristics such as the 80/20 training-validation split.
In generation of datasets for both systems, the computational resources required were a significant barrier.
Each MC simulation requires \O{10^7} s total wall-time.
We cannot simply exhaust the entire input space for either system -- regardless of the criticality.
Thus, a priori, we approached dataset generation by considering Latin hypercube sampling of the control device locations, with a minimum requirement of 100 sets each.
A total of 100 and 151 datasets were generated for the MGEP and MITR, respectively.
Using the 80/20 heuristic, we have a 80/20 and 121/30 split for the training and test datasets.

\begin{figure}
\includegraphics[width=\linewidth]{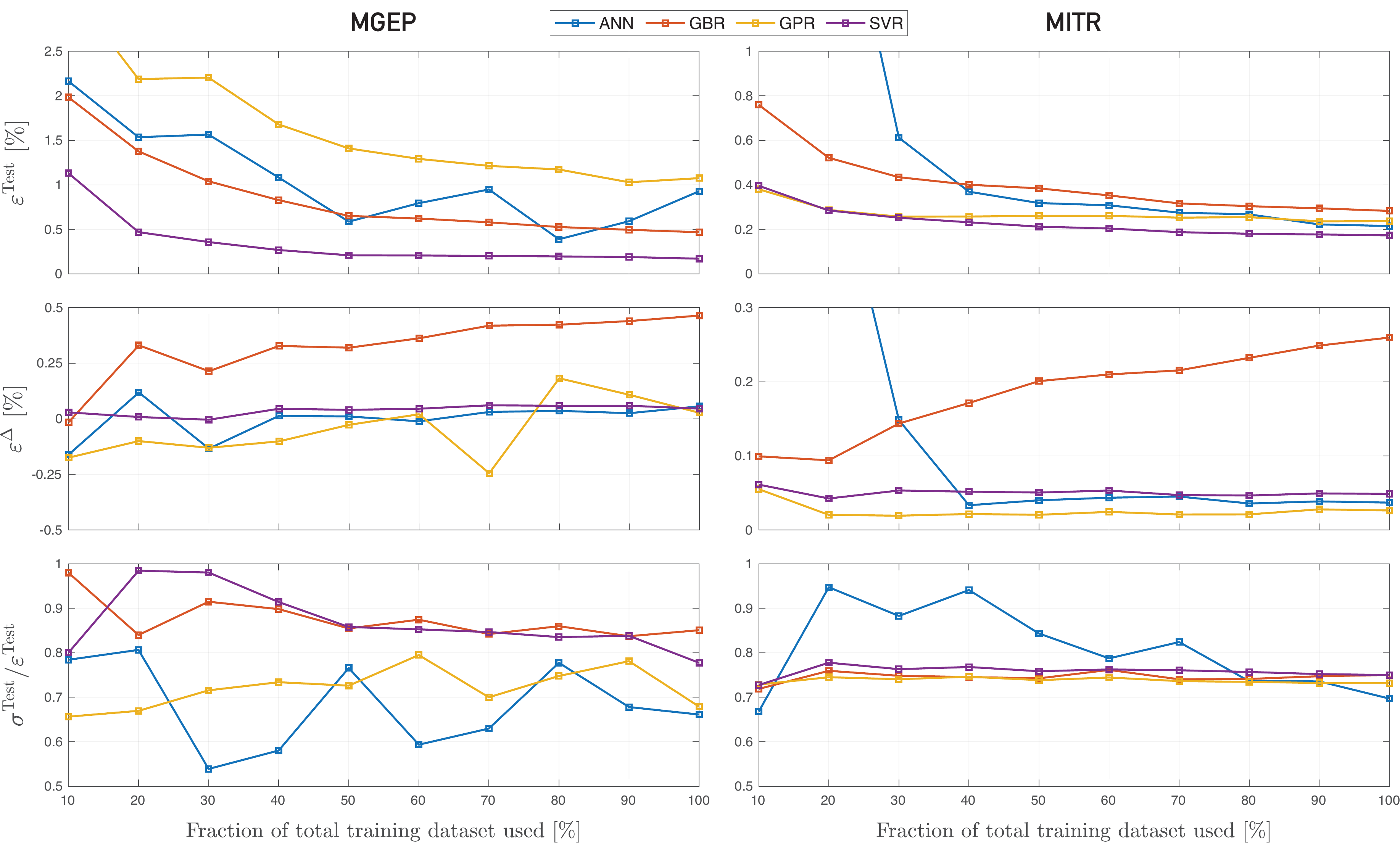}
\caption{Performance of algorithms with varying number of training datasets.
The domain-averaged test accuracy is presented in the top row.
The domain-averaged generalization error is presented in the middle row.
The domain-averaged precision, scaled by the test-error, is presented in the bottom row.
There were a total of 80 and 121 training datasets for the MGEP and MITR, respectively.
The test dataset is always kept constant, at 20 and 30 for the MGEP and MITR, respectively.
The optimal settings for each system and algorithm combination are used (tabulated in \cref{table:hypss}).}
\label{fig:crerror}
\end{figure}

The propagation of performance metrics with respect to total training dataset size, is presented in \cref{fig:crerror}.
First, the trends observed for the test accuracy, $\varepsilon^\mathrm{Test}$, are discussed.
For both systems, a general trend of exponential decay in test accuracy is noted.
The decay is sharpest for the MITR ANN model.
An interesting observation is that the GBR, GPR, and SVR trajectories do not overlap
Thus, given equivalent training datasets, SVR always outperforms GBR and GPR.
The ANN trajectory for the MGEP is erratic, indicating sensitivity to training \vs test set overlap.
Next, the trends observed for the generalization error, $\varepsilon^\Delta$, are discussed.
The GBR model suffers from overfitting as generalization error increases as dataset size increases.
As dataset size increases, the ANN model has marginal benefit \vs SVR.
Finally, the trends observed for the normalized precision, $\sigma^\mathrm{Test}/\varepsilon^\mathrm{Test}$, are discussed.
For the MITR, we note an almost equal propagation for the GBR, GPR and SVR models.
The ANN model has a greater variance at low dataset sizes, but eventually equalizes to other models.
For the MGEP, the variance for GBR and SVR are higher than that observed for the ANN and GPR at all dataset sizes.
However, we again observe that the trend for ANN is erratic.

The observations from probing training dataset size concludes in guidelines for creating empirical models for fission systems.
If training dataset size (or computational resources) is a constraint, the SVR algorithm is the optimal choice.
The ANN algorithm has marginal generalization benefits and moderate precision improvements over SVR at large dataset sizes.
Thus, the ANN algorithm is more data intensive than the SVR.
The GPR algorithm exhibits good generalization and precision, at the cost of poor accuracy.
The GBR algorithm is not recommended at any dataset size.

\section{Concluding Remarks}\label{sec:conclusions}

The accurate prediction of nuclear transport in fission systems is important in the design of nuclear power plants and characterization of their safety margin.
Modern safety analyses requires a multidimensional approach to optimize plant operation.
The state-of-the-art solutions of the neutron transport equation, \cref{eq:transport}, involve Monte Carlo simulations that are expensive and require \O{10^7} s total wall-time for a single solution.
Such an approach is feasible for offline calculations \eg to design the plant fuel loading pattern.
However, less computationally intensive methods are required if we want to embed high-fidelity model predictive controllers in autonomous control systems of the future.
To address this, we explore the capability of Artificial Neural Networks (ANNs), Gradient Boosting Regression (GBR), Gaussian Process Regression (GPR) and Support Vector Regression (SVR) algorithms to provide functional approximations of multidimensional neutron transport.
We generate empirical models for two fission systems that differ qualitatively: the \textit{subcritical} MIT Graphite Exponential Pile (MGEP) and the \textit{critical} MIT Research Reactor (MITR).
Monte Carlo (MC) simulations of both MGEP and MITR are used as training datasets.
To generate unique datasets, the positions of the control devices were varied asymmetrically using Latin hypercube sampling. 

In order to contrast the capability of each Machine Learning (ML) algorithm, a meta-learning procedure was implemented to systematically find optimal hyperparameter settings.
The optimal settings for each combination of nuclear system and ML algorithm are listed in \cref{table:hypss}.
The major findings of the optimization process are:
\begin{itemize}
\item The \textbf{optimal settings for the MGEP and MITR have a mix of similarities and differences}.
The similarities are generally in the category of hyperparameters that impact the model training process, \eg the adam optimizer for ANNs and the FMSE split criterion for GBR.
The differences were mostly in the structure of the model itself, \eg the shape and number of layers for ANNs, the maximum tree depth for GBR or the number of inducing points for GPR.
For all four ML algorithms considered, MGEP models required `larger' capacity than those of the MITR.
This finding was attributed to the lower input-output ratio of the MGEP model.
\item The improvement in training error across all hyperparameter sets was several orders of magnitude for the ANN, GPR and SVR models. 
GBR training was least sensitive to hyperparameter selection.
In observation of the bottom decile of hyperparameter sets, we found few coherent patterns to guarantee poor performance.
In other words, \textbf{manual selection of hyperparameters may lead to erratic and poor performance}, discouraging further use.
A systematic hyperparameter optimization procedure is essential to production of optimal ML models for fission systems.
\end{itemize}
The meta-learning procedure affirms that if we have a priori knowledge of the qualitative differences expected between systems, we should anticipate that separate optimal hyperparameter sets exist.
In other words, transfer learning \citep[Ch. 15]{goodfellow2016deep} between systems would not be applicable.
Next, the optimized ML models were contrasted on performance metrics such as accuracy, precision and generalization error, quantified by \cref{eq:acc,eq:std,eq:gen}, respectively.
The major findings of evaluating each optimal model are:
\begin{itemize}
\item For accuracy, the SVR model achieved a $\varepsilon^\mathrm{Test}$ of 0.17 \% for both MGEP and MITR.
The ANN model was next, achieving $\varepsilon^\mathrm{Test}$ of 0.37 \% and 0.22 \% for the MGEP and MITR.
For precision, the ANN and SVR models had similar $\sigma^\mathrm{Test}$ at 0.15 \% and 0.13 \% for the MITR.
For the MGEP, the corresponding values were 0.23 \% and 0.13 \%.
The ANN and SVR models had similar generalization error, $\varepsilon^\Delta\approx$ 0.05 \%.
For context, the $3\sigma$ power measurement uncertainty is 5.0 \% for the MITR \cite{dave2020thermal}, and the $3\sigma$ flux measurement uncertainty is 6.0 \% for the MGEP.
Therefore, on \textit{test} data, the \textbf{ANN or SVR empirical model performance is within experimental uncertainty} bounds.
The statement holds true even when considering local maxima rather than domain-averaged metrics.
The GPR model had poor accuracy and precision, yet exhibited the lowest generalization error for both systems.
The GBR model had moderate accuracy and precision, but very high generalization error.
\item Evaluating the spatial distribution of performance metrics reveals that \textbf{locations of poor performance coincide with locations at which perturbation of the neutron flux distribution are greatest}.
In general, ANN and SVR exhibit the most uniform spatial performance across all metrics considered.
This finding suggests that if we have a priori knowledge of a particular spatial region that is important for safety analysis, additional evaluation is required to assess if that region also experiences significant perturbation from control device movement.
\item The propagation of accuracy, generalization error and precision \vs training dataset size showed that the \textbf{SVR algorithm has a significant advantage when the availability of training data is low}.
The ANN algorithm for MITR has the most drastic improvement in accuracy and precision as dataset size increased.
The GBR algorithm suffered from overfitting for both systems, deteriorating as dataset size increased.
Therefore, if dataset availability is a constraint, the SVR algorithm is recommended over ANN.
\item The execution time for the ANN was the least, on the order of hundredths of a millisecond.
Both GBR and SVR required on the order of a millisecond for each evaluation.
The GPR model required greater than a millisecond for each evaluation.
The ANN performance supremacy stems from the lightweight formulation for multidimensional regression.
Importantly, both \textbf{ANN and SVR models achieve a greater than 7 order reduction in evaluation time} \vs a MC simulation.
\end{itemize}

Overall, SVR is recommended followed closely by ANN.
The ANN algorithm is recommended only if a sizable dataset is available.
GBR is not recommended for fission systems.
GPR can be a useful path to quantify posterior uncertainty through probabilistic modeling.
To further explore the objective of generating surrogate models for multidimensional regression of fission systems, there are several paths.
Physics-informed neural networks \citep{raissi2017physics} have gained traction as a physics-based approach -- however, it is unclear how macroscopic cross-sections, $\Sigma$ in \cref{eq:transport}, will be resolved in this context.
Another task is to contrast the performance of lower-order diffusion-based models (introduced in \cref{appeq:diff1,appeq:diff2}), \vs the empirical models generated from high-fidelity MC methods explored in this work. 

Efforts to address the objectives of this work resulted in the development of an open-source python package \pkg{NPSN} (\href{https://github.com/a-jd/npsn}{github.com/a-jd/npsn}).
The package abstracts the optimization and training across multiple ML algorithms to several lines of code, as presented in \cref{appendix:B}.
Our ongoing project is utilizing \pkg{NPSN} to generate empirical models that are deployed in a control system framework to autonomously control the MGEP.
The project is the first of its kind to embed ML controllers in an experimental demonstration of an autonomously controlled fission system.

\section*{CRediT authorship contribution statement}
\textbf{Akshay J. Dave}: Conceptualization, Methodology, Software, Writing – Original Draft, Visualization, Funding acquisition.
\textbf{Jiankai Yu}: MGEP Data curation.
\textbf{Jarod Wilson}: MITR Data curation, Writing - Review \& Editing.
\textbf{Bren Phillips}: Writing - Review \& Editing.
\textbf{Kaichao Sun}: MGEP and MITR Data curation, Writing - Review \& Editing, Supervision, Funding acquisition.
\textbf{Benoit Forget}: Writing – Review \& Editing, Supervision.

\section*{Acknowledgements}
This work is supported by US Department of Energy NEUP Award Number: {DE-NE0008872}.

\bibliographystyle{unsrtnat}
\bibliography{references}

\clearpage
\appendix
\section*{Appendices}
\section{Canonical Nuclear Systems}\label{appendix:A}

In order to demonstrate the qualitative differences between a subcritical and critical fission system, exact solutions for a canonical geometry with simplifying assumptions are derived.
We begin by considering the integro-differential neutron transport equation \citep[Ch. 4.II]{duderstadt1976nuclear},
\begin{multline}
\frac{1}{v}\frac{\partial}{\partial t}\phi(\mathbf{x},E,\mathbf{\Omega},t)+\mathbf{\Omega}\cdot\mathbf{\nabla}\phi(\mathbf{x},E,\mathbf{\Omega},t)+\Sigma_t(E)\phi(\mathbf{x},E,\mathbf{\Omega},t)=\\
\int_{4\pi}d\mathbf{\Omega}'\int_0^\infty dE'\Sigma_s(E'\rightarrow E,\mathbf{\Omega}'\rightarrow\mathbf{\Omega})\phi(\mathbf{x},E,\mathbf{\Omega},t)+s(\mathbf{x},E,\mathbf{\Omega},t)~,
\label{appeq:transport}
\end{multline}
where $v$ is the neutron speed (unit of length per time), $\phi$ is the angular neutron flux (unit of neutrons per unit area, energy, angular direction, and time), $\Sigma_t$ and $\Sigma_s$ are the neutron total and scattering macroscopic cross sections (unit of per unit length), and $s$ is the neutron source term (unit of neutrons per unit volume, energy, angular direction, and time).
Each term in \cref{appeq:transport} represents a specific physical process describing a rate of gain or loss of neutrons from $dVd\Omega dE$ about $(\mathbf{x},E,\mathbf{\Omega})$.
The first term on the left hand side accounts for time rate of change of neutrons.
The second term accounts for spatial diffusion of neutrons.
The third term accounts for total collisions  of neutrons.
The first term on the right hand side accounts for transport due to scattering of neutrons from energy $E'\rightarrow E$ and direction of flight $\mathbf{\Omega}'\rightarrow\mathbf{\Omega}$.
The second term is the neutron source term, which may due to fission reactions, emission from external neutron sources, or both.

Consider a steady-state, finite 1-D slab consisting of uniformly distributed fissionable medium, and a 1-group energy distribution.
Beginning from \cref{appeq:transport}, the steady-state neutron transport equation is,
\begin{equation}
\mathbf{\Omega}\cdot\mathbf{\nabla}\phi(\mathbf{x},E,\mathbf{\Omega})+\Sigma_t(E)\phi(\mathbf{x},E,\mathbf{\Omega})=
\int_{4\pi}d\mathbf{\Omega}'\int_0^\infty dE'\Sigma_s(E'\rightarrow E,\mathbf{\Omega}'\rightarrow\mathbf{\Omega})\phi(\mathbf{x},E,\mathbf{\Omega})+s(\mathbf{x},E,\mathbf{\Omega})~,
\label{appeq:transport2}
\end{equation}
dropping the first term and the independent variable $t$.
Next dependence on energy, $E$, is considered.
In fission systems, treatment of energy is a significant consideration as neutrons are `born' at high energies ($\approx\SI{2}{\mega\eV}$), and through scattering interactions with media, lose energy and are ultimately, absorbed at varying energy levels.
For thermal-spectrum reactors, absorption at low energies ($\approx\SI{1}{\eV}$) leads to fission.
However, fast-reactor designs rely on fission due to absorption at higher energies.
Therefore, appropriate energy group binning is important to accurately model a system.
To proceed with the 1-D slab proposed, a 1-group energy distribution is considered, dropping explicit dependence on energy,
\begin{equation}
\mathbf{\Omega\cdot\mathbf{\nabla}\phi(\mathbf{x},\mathbf{\Omega})}+\Sigma_t\phi(\mathbf{x},\mathbf{\Omega})=
\int_{4\pi}d\mathbf{\Omega}'\Sigma_s(\mathbf{\Omega}'\rightarrow\mathbf{\Omega})\phi(\mathbf{x},\mathbf{\Omega})+s(\mathbf{x},\mathbf{\Omega})~.
\label{appeq:transport3}
\end{equation}
Next the treatment on angular dependence is considered.
The direction of neutron flight is important in heterogeneous geometries where anisotropic neutron interactions play an important role \eg thermal backscattering at fuel boundaries.
In assuming that all interactions are isotropic, we can use the diffusion approximation.
Integrating over the surface of a sphere, the scalar neutron flux, scalar source, and neutron current is,
\begin{equation}
\phi(\mathbf{x})  = \int_{4\pi}d\Omega\phi(\mathbf{x},\mathbf{\Omega})~;~~
s(\mathbf{x})  = \int_{4\pi}d\Omega s(\mathbf{x},\mathbf{\Omega})~;~~
\mathbf{J}(\mathbf{x})  = \int_{4\pi}d\Omega\mathbf{\Omega}\phi(\mathbf{x},\mathbf{\Omega})~.
\label{appeq:def1}
\end{equation}
Using the terms defined in \cref{appeq:def1}, and integrating \cref{appeq:transport3} over a unit sphere,
\begin{equation}
\mathbf{\nabla\cdot J}(\mathbf{x})+\Sigma_a\phi(\mathbf{x})=s(\mathbf{x})~,
\end{equation}
In this context, the total macroscopic cross section is defined as the sum of the scattering and absorption interactions, $\Sigma_t=\Sigma_a+\Sigma_s$. The first term on the left hand side is simplified by Fick's law \cite[Ch. 4.IV.C]{duderstadt1976nuclear} for diffusion of neutrons,
\begin{equation}
\mathbf{J}(\mathbf{x}) = -D\mathbf{\nabla}\phi(\mathbf{x})~,
\label{appeq:diff1}
\end{equation}
where $D$ is a diffusion coefficient, quantifying how `freely' neutrons flow within a medium. 
Ficks law is analogous to Fourier's law for thermal conductivity.
The law is another simplifying assumption that is not appropriate for highly anisotropic conditions.
The resulting equation is known as the one-speed diffusion approximation equation,
\begin{equation}
D\nabla^2\phi(\mathbf{x}) - \Sigma_a\phi(\mathbf{x}) + s(\mathbf{x})=0~,
\label{appeq:diff2}
\end{equation}
and for the 1-D slab geometry,
\begin{equation}
D\frac{d^2}{dx^2}\phi(x) - \Sigma_a\phi(x) + s(x)=0~.
\end{equation}
Next, we assume that the neutron source, $s(x)=\nu\Sigma_f\phi(x)$, where $\nu$ is a constant to account for average neutrons born per fission, and $\Sigma_f$ is the macroscopic fission cross section,
\begin{equation}
\frac{d^2}{dx^2}\phi(x) + \frac{\nu\Sigma_f-\Sigma_a}{D}\phi(x)=0~.
\end{equation}
The infinite medium criticality, $k_\infty$, is defined by the ratio of neutrons born via fission \vs neutrons absorbed, \ie $k_\infty\equiv\nu\Sigma_f/\Sigma_a$.
The effective criticality, $k \equiv k_\infty P_\mathrm{NL}$, reduces by the probability of non-leaking, \ie neutrons staying within the system, $P_\mathrm{NL}$.
The diffusion length is defined as $L^2 \equiv D/\Sigma_a$.
With these definitions, the 1-D one-speed diffusion approximation for a uniformly distributed fissionable material is,
\begin{equation}
\frac{d^2}{dx^2}\phi(x) + \frac{k_\infty-1}{L^2}\phi(x)=0~.
\end{equation}
In subcritical systems, $k<1$ and an external source is required to sustain neutron populations.
In critical systems $k\gtrsim1$ and the fission reaction is self-sustaining (power reactors are designed to operate at $k$ slightly above 1).
Assuming $P_\mathrm{NL}\approx1$, for subcritical systems the solution of the second order ODE is of the form,
\begin{equation}
\phi(x) = A_1\exp(-x) + A_2\exp(x)~,
\label{eq:subsol}
\end{equation}
and similarly for critical systems,
\begin{equation}
\phi(x) = B_1\cos(Cx) - B_2\sin(Cx)~.
\label{eq:supsol}
\end{equation}
With suitably assigned boundary conditions, we can arrive at exact solutions for both systems.
The key takeaway is that \textit{solutions for both systems differ qualitatively}.
The spatial distribution is exponential for subcritical systems, and sinusoidal for critical. 
A priori, we expect that the the empirical models produced in this work for the MIT Graphite Exponential Pile (subcritical) and the MIT Research Reactor (critical) systems, will differ qualitatively.
A partial objective of this work is to investigate and disseminate these differences.

\section{NPSN Interface}\label{appendix:B}

To address the objectives of this work and facilitate, an open-source python package, \pkg{NPSN} (\href{https://github.com/a-jd/npsn}{github.com/a-jd/npsn}), was developed.
The layout of the package is summarized in \cref{fig:npsn}.
Installation instructions for the package are available on the website.
The interface used to generate empirical models from training data is presented below.
On line 1, the package is imported into the python environment.
On line 4, the location of the comma separated value (CSV) files used to train the models is input.
On line 6, a project name is input. The project name is used as a suffix for output files.
On line 9, the algorithm chosen is defined (currently valid values are ANN, GBR, GPR, and SVR).
On line 11 and 13, the values of $c$ and $e,~n$ in \cref{eq:map} are defined, respectively.
On line 16 and 18, the training command is sent. If the number of evaluations requested is greater than 1, hyperparameter optimization is performed.
Lastly, on line 20, the performance metrics defined in \cref{eq:acc,eq:std} are output as CSV files.

\begin{lstlisting}
import npsn

# Define dataset directory
data_dir = "/some/data_location"
# Define project name (for output file label)
proj_nm = "npsn_surrogate"

# Define algorithm type to be used
algo_nm = "ANN"
# Define number of control devices
n_x = 6
# Define nodalization of power distribution
n_y = (16, 22)  # (e, n)

# Train without optimization
npsn.train(proj_nm, algo_nm, data_dir, n_x, n_y)
# OR with optimization
npsn.train(proj_nm, algo_nm, data_dir, n_x, n_y, max_evals=100)
# Post-process to quantify error (CSV file output)
npsn.post(proj_nm)
\end{lstlisting}

\end{document}